\documentstyle[aps,floats,epsf,psfig,twocolumn]{revtex}
\tighten
\begin{document}
\draft 
\title{Theory of Electronic Raman Scattering in Nearly Antiferromagnetic Fermi
Liquids}
\author{T.P. Devereaux$^{1}$ and A.P. Kampf$^{2}$}
\address{$^{1}$
Department of Physics, George Washington University, Washington, DC 20052}
\address{$^{2}$
Theoretische Physik III, Elektronische Korrelationen und Magnetismus, \\
Institut f\"ur Physik, Universit\"at Augsburg, 86135 Augsburg, Germany}
\address{~
\parbox{14cm}{\rm 
\medskip
A theory of electronic Raman scattering in nearly antiferromagnetic Fermi 
liquids is constructed using the phenomenological electron-electron 
interaction introduced by Millis, Monien, and Pines. The role of "hot spots"
and their resulting signatures in the channel dependent Raman spectra is
highlighted, and different scaling regimes are addressed. The effects of
self consistency and vertex corrections are compared to a perturbative
treatment of the interaction. The theory is then 
compared to Raman spectra taken in the normal state of 
cuprate superconductors, and it is shown that many 
features of the
symmetry dependent spectra can be explained by the theory. Remaining
questions are addressed.
}}
\maketitle
%\pacs{PACS numbers: 74.25.Jb, 71.27.+a, 61.16.Ch}

\narrowtext
\section{Introduction.}

Recent results from angle resolved photo-emission (ARPES) studies have shown
that the planar quasiparticle (qp) dynamics in the normal state of
high temperature superconductors are extremely anisotropic\cite{marshall}. 
For optimally doped samples with the highest T$_{c}$'s, the qp
spectral function 
observed for Fermi surface crossings along the $(\pi,0)-(\pi,\pi)$
direction in the Brillouin zone (BZ) is smeared by strong
scattering compared to the observed spectrum along the
$(0,0)-(\pi,\pi)$ direction.
This anisotropy becomes more pronounced as the materials are 
further underdoped with concomitantly lower transition temperatures
into the superconducting state\cite{pgap}. 
These effects have been ascribed to a strongly momentum dependent qp
self energy $\Sigma({\bf k},\omega)$ whose imaginary part is 
largest for ${\bf k}$ near
the BZ axes and yields a broadly incoherent spectral function.
Understanding the nature and origin of
this anisotropy could lead to a better understanding of the complex
dynamics which cause the normal state of the cuprates to be
very far away from being ``normal''. 

Many theoretical efforts have been conducted to understand this planar 
anisotropy by calculating $\Sigma({\bf k},\omega)$ for
various types of interactions and via several methods. 
These efforts usually have
focused on understanding the results from photo-emission. Since ARPES
is a single particle probe, the spectral function
can be compared directly with experiment and provides direct
contact to the self energy around the BZ. However, most other probes 
(e.g. penetration depth, conductivity, tunneling density of states, etc.)
involve
averages around the BZ and thus are not so sensitive to the effects of
anisotropic self energies. Electronic Raman scattering is an exception. 
As demonstrated in the superconducting state\cite{ijmpb}, electronic
Raman scattering is a two particle probe which
has proven to be a useful tool to understand qp dynamics on selected 
regions of the BZ. The matrix elements for 
Raman scattering can be tuned by orienting
incoming and outgoing photon polarizations, thereby projecting out 
different regions of the Fermi surface. In Ref. \cite{ijmpb} it
was shown how this tuning could be used to provide a detailed probe
of the momentum dependence of the superconducting energy gap.

In this regard, Raman scattering can be considered a complementary probe to 
ARPES. Raman measurements on several systems have shown a remarkable 
{\it channel} dependence in the normal state for the optimally- and under-doped
cuprates in a variety of systems\cite{ubcdata,rhoraman,hackl,psuedo,psuedo2}. 
For scattering geometries of 
$B_{1g}$ symmetry, an almost\cite{psuedo2} 
temperature independent Raman spectrum 
has been observed up to frequencies of a few $J$,
while for $B_{2g}$ orientations the spectra show
a temperature dependence that tracks that of the DC resistivity
\cite{rhoraman,psuedo}.
Moreover, the overall magnitude of the electronic continuum is observed
to be relatively independent of doping for the $B_{2g}$ channel, but
decreases dramatically with underdoping for the $B_{1g}$ channel, with
spectral weight transferring out to two-magnon energies\cite{psuedo,psuedo2}.
One is to conclude that the qps near $(\pi,0)$ which are 
probed by the $B_{1g}$ channel are particularly incoherent and become
more incoherent with underdoping, in agreement
with the results from ARPES. Coupled to 
the flatness of the bands there, these qps are likely not to 
participate in transport\cite{im}. 
Currently there is no theoretical understanding of Raman scattering in the
normal state of the cuprates, as all previous theories failed
to predict large differences between these two channels at low frequencies.
This is due to the neglect of strongly anisotropic electron interactions.
It is the purpose of this paper to address these issues.

It is widely believed that strong antiferromagnetic (AF) correlations are an
important ingredient needed to describe the unusual properties found in
the normal state of the cuprate superconductors and provides
the origin of the strongly anisotropic dynamics in the cuprates. Large
qp scattering for momentum transfers near the AF
wave-vector ${\bf Q}={\pi/a,\pi/a}$ 
have been invoked to explain the photo-emission 
results\cite{zxrs,altmann}. Therefore we have studied 
the phenomenological Nearly Antiferromagnetic Fermi Liquid (NAFL) model 
\cite{MMP} as an example of highly
anisotropic scattering to investigate the two-particle Raman
response. The NAFL model has at its center a strong magnetic interaction
between planar qps which is peaked at or near ${\bf Q}$ and yields
anisotropic self energies. This model has been widely
applied to investigate both spin and charge properties\cite{rmp,ps}. 
Our results show that the Raman response is an ideal probe
to clarify the nature of qp dynamics on regions of the
Fermi surface. 

The outline of the paper is as follows: Section II reviews and outlines
a generalized approach to calculate non-resonant Raman scattering in
correlated electron systems and introduces the model interaction. Section
III investigates the NAFL model
using simple perturbation theory. The scattering rate for qps
on different regions of the Fermi surface is calculated and the subsequent
Raman response is obtained. Section IV compares the results of the
calculations in Section III to data on overdoped
Bi$_{2}$Sr$_{2}$CaCu$_{2}$O$_{8+\delta}$ (Bi-2212). Section V investigates the
role of a self consistent treatment of the interaction and the role
of vertex corrections. Finally Section VI summarizes our results and
states our conclusions.

\section{Formalism.}

The intensity of inelastically
scattered light can be written in terms 
of a differential photon scattering cross section as 
\begin{eqnarray}
\frac{\partial^2\sigma}{\partial\omega\partial\Omega}&=&
\frac{\omega_{\rm S}}{\omega_{\rm I}}\ r_0^2\ S_{\gamma\gamma}({\bf q},\omega),
\nonumber \\
S_{\gamma\gamma}({\bf q},\omega)&=&-\frac{1}{\pi}
\left[1+n(\omega)\right]{\rm Im}\ \chi_{\gamma\gamma}({\bf q},\omega).
\label{one}
\end{eqnarray}
Here $r_0=e^2/mc^2$ is the Thompson radius, $\omega_{\rm I}$, $\omega_{\rm S}$
are the frequency of the incoming and scattered photon, respectively, and
we have set $\hbar=k_{B}=1$.
$S_{\gamma\gamma}$ is the generalized structure function, which is related to 
the imaginary part to the Raman response function $\chi_{\gamma\gamma}$
through the fluctuation--dissipation theorem, the second part of Eq. (1).
$n(\omega)$ is the Bose--Einstein distribution function. 
The Raman response measures ``effective density'' fluctuations,
\begin{equation}
\chi_{\gamma\gamma}(i\omega)=\int_{0}^{1/T} d\tau\, e^{-i\omega\tau}
\langle T_{\tau}[\tilde\rho(\tau)\tilde\rho(0)]\rangle,
\label{two}
\end{equation}  
with $T_{\tau}$ the time-ordering operator and the imaginary part is
obtained by analytic continuation, $i\omega \rightarrow \omega +i0^+$. 

The Raman intensity can be represented as a scattering off
an effective charge density\cite{kandd}
\begin{equation}
\tilde\rho=\sum_{{\bf k},\sigma}\gamma({\bf k};\omega_{I},\omega_{S})
c^{\dagger}_{\sigma}({\bf k})c_{\sigma}({\bf k}),
\label{three}
\end{equation}
where $\sigma$ is the spin index, $\gamma({\bf k};\omega_{I},\omega_{S})$ 
is the Raman scattering 
amplitude, and $\omega_{I,S}$ is the frequency of the incident, scattered
photon, respectively. 
The effective charge density is a nonconserving quantity in 
contrast to the real charge\cite{kz}. Here the Raman vertices $\gamma$
are related to the incident and scattering photon
polarization vectors ${\bf e}^{I,S}$, resulting from a coupling of
both the charge 
current and the charge density to the vector potential. 
In general the Raman vertex
depends non-trivially on both the incident and scattered photon
frequencies.
According to Abrikosov and Genkin\cite{ag}, if the energy of the incident
and scattered frequencies $\omega_{I},\omega_{S}$ are negligible
compared to the energy band gap, $\gamma$
can be expressed in terms of the curvature of the bands and the incident
and scattered photon polarization vectors ${\bf e}^{I,S}$ as
\begin{equation}
\lim_{\omega_{I},\omega_{S} \rightarrow 0}
\gamma({\bf k};\omega_{I},\omega_{S})
=\sum_{\mu,\nu} e^{I}_{\mu}
{\partial^{2}\epsilon({\bf k})\over{\partial k_{\mu}
\partial k_{\nu}}}e^{S}_{\nu},
\label{four}
\end{equation}
where terms of the order of $1-\omega_{S}/\omega_{I}$ are dropped.
This expression is valid if the incoming laser light cannot excite
direct band-band transitions. However one might question the 
appropriateness of this approximation for the cuprates given
that typical incoming laser frequencies are on the order of 2 eV.

An alternative approach is based on the experimental observation that
the electronic continua in the metallic normal state for a
wide range of cuprate materials depend only mildly on the incoming laser
frequency.
Since the polarization orientations transform as various
elements of the point group of the crystal, one can use symmetry to classify
the scattering amplitude, viz.
\begin{equation}
\gamma({\bf k};\omega_{I},\omega_{S})=\sum_{L}\gamma_{L}(\omega_{I},\omega_{S})
\Phi_{L}({\bf k}),
\label{five}
\end{equation}
where $\Phi_{L}({\bf k})$ are either Brillouin zone (BZH, orthogonal 
over the entire Brillouin zone) or Fermi surface (FSH, orthogonal
on the Fermi surface only) harmonics which transform
according to point group transformations of the crystal\cite{allen}. 
Representing the magnitude but not the 
${\bf k}$-dependence of scattering, the prefactors
can be approximated to be frequency independent and taken as model constants 
to fit absolute intensities. 

Thus we have simplified the many-band problem in terms
of symmetry components which can be related to charge degrees of freedom
over portions of the BZ. While sacrificing information
pertaining to overall intensities, we have gained the ability to probe
and compare excitations on different regions of the BZ based
solely on symmetry classifications. This can be illustrated
by considering the various experimentally accessible polarization orientations.

Using an $x,y$ coordinate system locked to the CuO$_{2}$ planes, incident and 
scattered light polarizations aligned
along $\hat x+\hat y, \hat x-\hat y$ for example
transform according to $B_{1g}$ symmetry, and thus 
\begin{equation}
\Phi_{B_{1g}}({\bf k})=\cos(k_{x}a)-\cos(k_{y}a) + \dots,
\label{six}
\end{equation}
where $\dots$ are higher order BZH.
Likewise, ${\bf e}^{I,S}$ aligned
along $\hat x, \hat y$ transforms as $B_{2g}$:
\begin{equation}
\Phi_{B_{2g}}({\bf k})= \sin(k_{x}a)\sin(k_{y}a) + \dots.
\label{seven}
\end{equation}
The $A_{1g}$ basis function is
\begin{eqnarray}
&&\Phi_{A_{1g}}({\bf k})=
a_{0} + a_{2}[\cos(k_{x}a)+\cos(k_{y}a)] \\
&&+ a_{4}\cos(k_{x}a)\cos(k_{y}a)
+a_{6}[\cos(2k_{x}a)+\cos(2k_{y}a)] +\dots, 
\label{eight}
\nonumber
\end{eqnarray}
with the expansion parameters $a_{i}$ determined via a fitting procedure
with experiment. The $A_{1g}$ response is not directly accessible
from experiments and must be obtained by subtracting several combinations
of the response for various polarization orientations. 

By considering the ${\bf k}-$dependence of the basis functions, it is clear
that the $B_{1g}$ part of the spectra probes light scattering
events along the $k_{x}$ or $k_{y}$ axes, $B_{2g}$ probes the diagonals,
and $A_{1g}$ is a weighted average over the entire Brillouin zone. In 
this manner information about the momentum dependence of the 
qp scattering rate can be obtained. Since we are interested in probing
anisotropy effects we will focus on $B_{1g}$ and $B_{2g}$ and
not consider the $A_{1g}$ channel.

Since the momentum transfer by the photon is negligible compared to
the momentum of the electrons in metals, we are interested in the
${\bf q}=0$ Raman response. With the approximations above, the
gauge invariant Raman response is of the form
\begin{eqnarray}
&&\chi_{\gamma\gamma}({\bf q=0},i\Omega)= 
\label{nine}
\\
&&-{T\over N}\sum_{i\omega}\sum_{\bf k}
\gamma({\bf k}) G({\bf k},i\omega) G({\bf k},i\omega+i\Omega)
\tilde\gamma({\bf k},i\omega,i\omega+i\Omega),
\nonumber
\end{eqnarray}
with the renormalized vertex obeying a Bethe-Salpeter equation
$\tilde\gamma$
\begin{eqnarray}
&&\tilde\gamma({\bf k},i\omega,i\omega+i\Omega)=\gamma({\bf k})+
{T\over N}\sum_{i\omega^{\prime}}\sum_{\bf k^{\prime}}V({\bf k-k^{\prime}},
i\omega-i\omega^{\prime}) \nonumber \\
&&\hskip0.5cm\times G({\bf k^{\prime}},i\omega^{\prime})
G({\bf k^{\prime}},i\omega^{\prime}+i\omega)\tilde\gamma({\bf k^{\prime}},
i\omega^{\prime},i\omega^{\prime}+i\Omega).
\label{ten}
\end{eqnarray}
Here $V$ is the effective
qp interaction in the particle-hole channel to be discussed below, and
$N$ is the number of sites.
In the absence of vertex corrections to the Raman vertex $\gamma({\bf k})$, 
the Raman response is given by
\begin{eqnarray}
\chi^{\prime\prime}_{\gamma\gamma}({\bf q}={\bf 0},\Omega)=
&&{2\over{N}}\sum_{\bf k}\gamma^{2}({\bf k})
\int {{\rm d}\omega\over{\pi}}[f(\omega)-f(\omega+\Omega)]\nonumber \\
&&\times G^{\prime\prime}({\bf k},\omega)
G^{\prime\prime}({\bf k},\omega+\Omega).
\label{eleven}
\end{eqnarray}
The Green's function is given in terms of the self energy
$\Sigma$ by Dyson's equation,
$G({\bf k},\omega)=[\omega-\varepsilon({\bf k})-\Sigma({\bf k},
\omega)]^{-1}$. We use a 2-D tight binding band structure
$\varepsilon({\bf k})=-2t[\cos(k_{x}a)+\cos(k_{y}a)]
+4t^{\prime}\cos(k_{x}a)\cos(k_{y}a)-\mu$, and  
from now on we set the lattice constant $a=1$. 

\section{Perturbation Theory and Role of Scattering.}

\subsection{Other models.}

Before we introduce the NAFL portion of the calculation, we start by
considering the simple form for the Raman response given in Eq.
(\ref{eleven}). We first consider non-interacting electrons.
In that case,
the imaginary part of the Green's functions are simply $\delta$
functions, and thus the $q\rightarrow 0$ Raman response is given 
essentially by the Lindhard function weighted by the Raman vertex.
Since for $q=0$ there is no phase space available to create electron-hole
pairs at low energies, this limit would yield no Raman cross section.
Thereby, it is crucial to have a non-zero self energy in order observe
inelastic light scattering at all in this limit.

Zawadowski and Cardona considered free electrons in the presence of
a potential due to impurities\cite{zawa}. Using Eq. (\ref{nine} \&
\ref{ten}) and
assuming mainly isotropic impurity scattering, $\Sigma({\bf k},\omega)
=-i/\tau_{L=0}$, as well as a separable interaction diagonal in
Fermi surface harmonics $\phi_{L}$,
$V({\bf k-k^{\prime}})=V_{\bf k,k^{\prime}}=\sum_{L}V_{LL}\phi_{L}({\bf k})
\phi_{L}^{*}({\bf k^{\prime}})$, the resulting cross
section can be written in a Lorentzian form:
\begin{equation}
\chi_{LL}^{\prime\prime}(\Omega)=2N_{F}\gamma_{L}^{2} 
{\Omega \tau^{*}_{L}\over{1+(\Omega\tau^{*}_{L})^{2}}},
\label{twelve}
\end{equation}
where a flat band with density of states per spin at the Fermi level 
$N_{F}$ has been assumed, and $1/\tau^{*}_{L}=1/\tau_{L=0}-1/\tau_{L}$
is the impurity scattering rate in channel $L$ reduced by vertex
corrections. Finally $\gamma_{L}$
is the prefactor of the Raman vertex in channel $L$ (see Eq. \ref{five}).
While impurities do yield a channel dependent response if the $V_{LL}$
are different for different $L$, since the
spectra have a peak at the scattering rate and then
fall off at large $\omega$, it cannot
fit the large flat continuum observed in Raman experiments extending
up to the scale of a few eVs \cite{ubcdata,rhoraman,hackl,psuedo,psuedo2}.

A form similar to Eq. (\ref{twelve}) has been obtained for the case
of qp scattering in a nested Fermi liquid\cite{nfl}. The
calculated spectra have the same form as (\ref{twelve}) except that
a frequency and temperature dependent scattering rate is introduced
which is of the form $1/\tau^{*}_{L}=max\{\beta_{L} T, \alpha_{L}\omega\}$,
with channel dependent prefactors $\beta,\alpha$. This form for the
response gives qualitatively good agreement with the observed 
spectra\cite{nfl},
but does not yield appreciably large differences between response
functions for different channels. In principle features of the
qp interaction due to nesting, which is pronounced for
particular momentum transfers near the nesting wavevector, may well account for
the observed scattering anisotropy. However, calculations have not been
performed for model Fermi surfaces.

\subsection{NAFL Theory.}

Therefore, we now consider a form for the qp interaction
due to an effective spin-fermion model interaction\cite{MMP}
\begin{equation}
V({\bf q},\Omega)=g^{2}{\alpha\xi^{2}\over{1+({\bf q-Q})^{2}\xi^{2}-i\Omega/
\omega_{sf}}}.
\label{thirteen}
\end{equation}
This interaction forms the basis of the NAFL theory, and has been derived 
by considering the interaction of electrons to external spin degrees of 
freedom\cite{chub,rmp} and decribes at low temperatures and several dopings 
the effective interaction calculated via FLEX\cite{altmann} and 
Quantum Monte Carlo\cite{bulut} treatments of the 2D Hubbard model. 
Recently the optical conductivity was explored within 
this model and was found to be in relatively good agreement with 
experiments\cite{ps}. Our efforts complement this study and 
investigate the consequences of Eq. (\ref{thirteen}) in greater detail.

The self energy at lowest order is given by
\begin{equation}
\Sigma({\bf k},i\omega)=-{T\over{N}}\sum_{i\omega^{\prime}}\sum_{\bf p}
V({\bf k-p},i\omega-i\omega^{\prime})G({\bf p},i\omega^{\prime}),
\label{fourteen}
\end{equation}
with $G$ the single particle Green's function. 
Here $\omega_{sf}$ and $\xi$ 
are the phenomenological temperature dependent spin fluctuation energy scale 
and the correlation length, respectively, which have been determined via fits 
to magnetic response data\cite{rmp}. For optimally doped and underdoped
systems, the strong AF correlations produce three magnetic regions in
the normal state. In the phase diagram constructed in
Ref. \cite{rmp}, $\omega_{sf}$ and $\xi$
obey certain 
relations depending on temperature and doping regimes. 
For each scaling regime, $\omega_{sf}\xi^{z}=$ constant (i.e. 
temperature independent), with $z$ the dynamical critical exponent.
For the $z=2$ regime the spin correlations are governed by temperature driven
fluctuations at a temperature above $T_{cr}$. As the temperature is reduced
below $T_{cr}$ the $z=1$ or pseudo-scaling regime is reached where 
the spin correlations are strong enough 
to lead to changes from the classical mean field $z=2$ regime. 
$T_{cr}$ is taken to be doping dependent and
rises with underdoping. Therefore the $z=1$ region plays a more dominant
role in the phase diagram than $z=2$ for underdoped systems while for
appreciably overdoped systems, the $z=1$ region vanishes.
At lower temperatures still and away from overdoped systems a crossover to a
pseudogap state occurs at $T^{*}$. However, we will not concentrate 
on this region and defer a discussion of pseudogap effects to the last
section.

As in Ref. \cite{ps}, we use the following approximations:
(i) we neglect the real part of the self energy, (ii) evaluate 
the Green's function at lowest order in the coupling constant $g^{2}$, (iii) 
neglect all vertex corrections for the Raman
vertex, and (iv) momentum sums are replaced by $\sum_{\bf k}=\int{\rm d}
(\Omega_{\bf k}/\mid {\bf v_{\bf k}}\mid)\int d\varepsilon({\bf k})$.
While (iv) does not crucially affect the results, approximations (i-iii) --
while simplifying the calculations -- considerably miss important qp
renormalizations at larger values of the coupling. Therefore we 
expect that these calculations would be most appropriate to describe the 
spectra taken on over-doped cuprate superconductors. 

\subsubsection{Quasiparticle scattering rate.}

\begin{figure}
\psfig{file=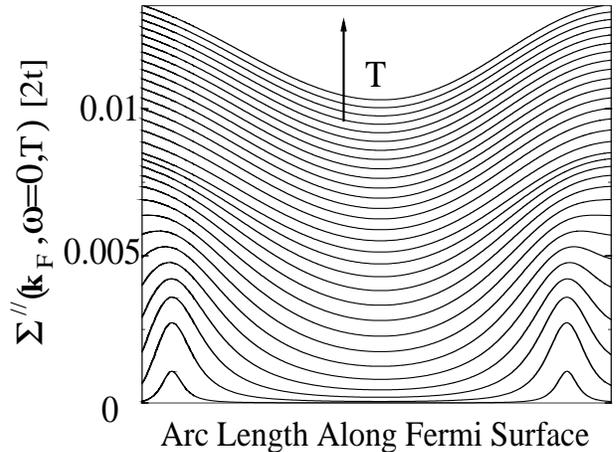,height=6.cm,width=8.cm,angle=0}
\caption[]{Quasiparticle scattering rate around a section of the Fermi 
surface for different temperatures in the NAFL model. Here we have taken
$\omega_{sf}/2t=0.00876+6.14T/2t$ for all $T$ and
$\omega_{sf}\xi/2t=0.1058$ for $T/2t<0.0273\, (z=1)$,
and then taken $\omega_{sf}\xi^{2}/2t=0.0635$ for $0.6>T/2t>0.0273\, 
(z=2)$.}
\label{fig1}
\end{figure} 

We first start by consider the effective qp scattering rate around the
Fermi surface. In Fig. \ref{fig1} we show the qp scattering rate, defined
as the imaginary part of the self energy $\Sigma({\bf k},\omega=0,T)$,
around the Fermi surface as a function of temperature $T$. Here 
the effective coupling constant is taken as $\alpha g^{2}/2t=6.25$. 
As a result of our perturbative approach, the magnitude of the scattering 
rate scales directly with $\alpha g^{2}$. The band structure parameters are 
chosen to be most applicable to slightly overdoped systems: 
$t^{\prime}/t=0.45$ and a filling $<n>=0.8$. The other parameters of the 
theory are chosen to provide a smooth link from the $z=2$ to the $z=1$ 
scaling regimes, as defined in the figure caption. 

Fig. \ref{fig1} shows that the scattering rate is extremely anisotropic
around the Fermi surface, having ``hot regions'' (near the zone axes)
where the qp scattering rate is peaked, and ``cold regions'' (near
the zone diagonals) where the rate is smallest. 
The ``hot spots'' are those
regions of the Fermi surface which can be connected
by ${\bf Q}$. These
regions are close to the zone axes for the given band structure, and the 
$B_{1g}$ geometry most effectively probes these hot spots while the
$B_{2g}$ geometry probes along the zone diagonals and therefore sees ``colder''
qps.  
The difference between 
the scattering at the ``hot'' and ``cold'' regions is smallest at
high temperatures, and is due largely to the enhanced density of states
near the zone axes. As the temperature cools the ``hot spots''
become more developed and more strongly peaked 
as the interaction becomes sharply
peaked for momentum transfers near ${\bf Q}$.

\begin{figure}
\psfig{file=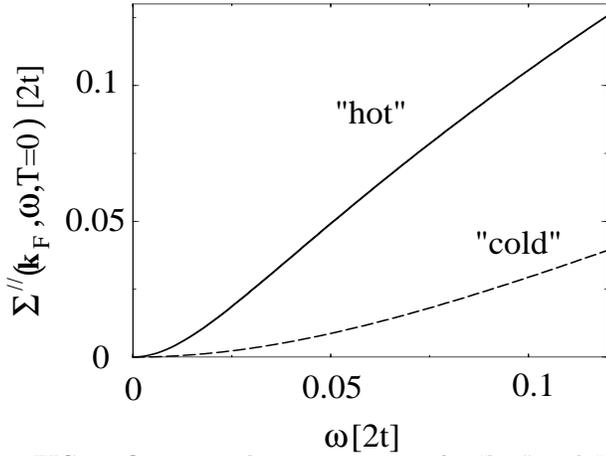,height=6.cm,width=8.cm,angle=0}
\caption[]{Quasiparticle scattering rate for ``hot'' and ``cold''
spots on the Fermi surface as a function of frequency for $T=0$.
The ``hot'' spot corresponds to ${\bf k_{F}}=(3.026,0.4817)$
while the ``cold'' spot corresponds to ${\bf k_{F}}=(1.107,1.107)$.
Here the $z=1$
parameters as given in the caption of in Fig. \ref{fig1} are used.}
\label{fig2}
\end{figure} 

Next we consider the frequency dependence of the scattering rate away from 
the Fermi level. This is shown in Fig. \ref{fig2}, where the qp scattering 
rate at $T=0$ is plotted for both ``hot'' and ``cold'' qps, as defined in the 
figure caption. Both scattering rates vary as $\omega^{2}$ at small $\omega$, 
characteristic of a Fermi liquid, but cross over to a behavior which is
effectively linear in $\omega$ at larger frequencies. For the ``hot''
qps this crossover happens at smaller values of $\omega$ (0.025, in
units of $2t$) than for the ``cold'' qps (0.092). Moreover, the scattering
is once again larger for the ``hot'' spots than for the ``cold''.
Similar results have been obtained in Ref. \cite{ps,rmp}, and the
reader is directed to that reference for further details. From the
point of view of Raman scattering, we would expect that the different
magnitude and frequency/temperature 
dependence of the scattering rate around the Fermi surface will
be manifest in the channel dependent Raman response. 
 
\subsubsection{Channel dependent Raman response.}

Considering the Raman response, we see that the $B_{1g}$ part
of the spectra probes the ``hot regions'' of large qp scattering
while $B_{2g}$ probes the ``cold regions''. This means that the
$B_{1g}$ response should probe less coherent qps than the
$B_{2g}$ response, which should be reflected in the temperature
dependence and general lineshape of the Raman continuum.

\begin{figure}
\psfig{file=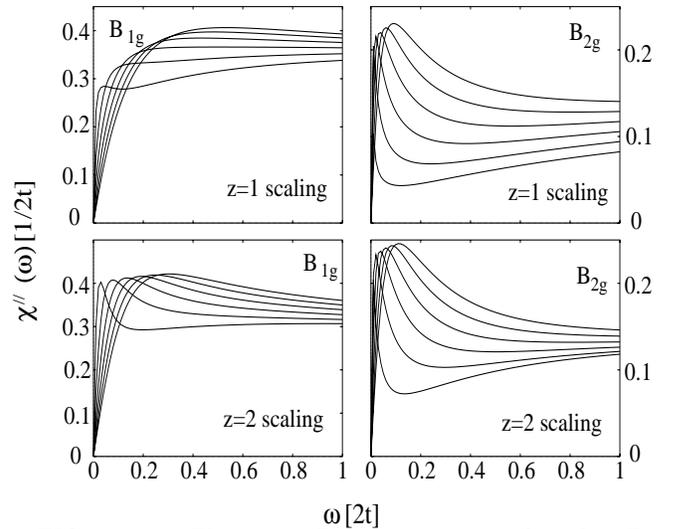,height=7.cm,width=8.5cm,angle=0}
\caption[]{
Electronic Raman response for the $B_{1g}$ and $B_{2g}$ channels evaluated at 
different temperatures ($T/2t=0.01,0.02,...,0.06$ from top to bottom) for the
$z=1$ and the $z=2$ scaling regimes of NAFL theory. Here we have set 
$b_1=b_2=1$, as defined in the text.}
\label{fig3}
\end{figure}

Within our perturbative approach we present in Fig. \ref{fig3} the results 
for the $B_{1g}$ and $B_{2g}$ Raman response functions for both $z=1$ and
$z=2$ scaling regimes. The parameters we have used for both scaling regimes 
are $g=1$eV, $\alpha=3.1$ states/eV, $t=250$meV,
$t^{\prime}/t=0.45$, filling $<n>=0.8$, $\gamma_{B_{1g}}({\bf k})=
b_{1}[\cos(k_{x})-\cos(k_{y})]$, $\gamma_{B_{2g}}({\bf k}) =
b_{2}\sin(k_{x})\sin(k_{y})$. In addition, for the $z=1$ scaling regime we
have used $\omega_{sf}\xi=50$meV and $\xi^{-1}=0.1+4.64 T/2t$, while 
for the $z=2$ scaling regime, $\omega_{sf}\xi^{2}=60$meV and 
$\omega_{sf}/2t=0.0237+0.55 T/2t$. These parameters are similar to those used
in Ref. \cite{ps} to describe the Hall conductivity data in 
YBa$_{2}$Cu$_{3}$O$_{7}$. Here the absolute magnitude for the scattering is 
arbitrary and determined by the dimensionless coefficients $b_{1}$ and 
$b_{2}$, which are set by fitting to the data. This has no effect on the 
frequency dependent lineshapes.

The spectra for both scaling regimes share several features. First, the flat 
continuum at high frequencies which is present in the Raman data from all
cuprate superconductors is reproduced by the theory. This is a consequence 
of a scattering rate $\Sigma^{\prime\prime}(\omega)$ which is effectively 
linearly dependent on $\omega$ at frequency scales larger than $\omega_{sf}$, 
as shown in Fig. \ref{fig2}. The spectra rise linearly with $\omega$ with a 
smaller slope for large scattering than for weaker scattering.
More importantly, the Raman response is 
different for the two scattering geometries as a consequence of
the strong anisotropy of the scattering rate. 
This is due to the relative magnitude and frequency dependence of
the scattering rate. As shown in Figs. \ref{fig1} and 
\ref{fig2}, the scattering rate probed
in the $B_{1g}$ channel is larger than $B_{2g}$ for all frequencies and
temperatures and becomes effectively linear in frequency at
a much smaller frequency than that probed by the $B_{2g}$ channel. 
Thus the position of the peak of the spectra in 
the $B_{1g}$ channel is at a larger frequency in all figures compared to the 
$B_{2g}$ channel, and the $B_{1g}$ spectra is overall flatter than
the $B_{2g}$ spectra. 

\begin{figure}
\psfig{file=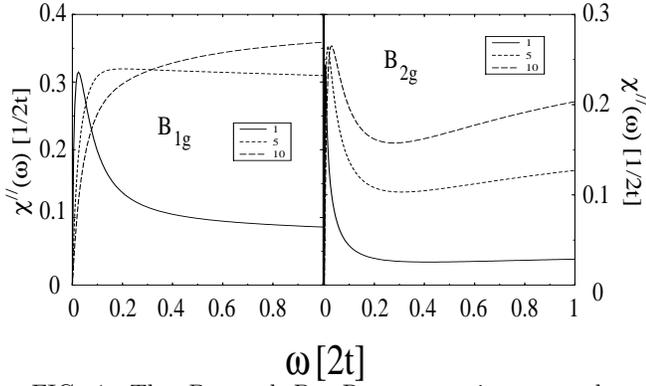,height=5.cm,width=8.5cm,angle=0}
\caption[]{The $B_{1g}$ and $B_{2g}$ Raman continua are plotted for
$T/2t=0.03$ for values of the effective coupling constant $g_{eff}=
1,5,10$, as defined in the text. The $z=1$ parameters were used
for $\omega_{sf}$ and $\xi$ as in Fig. \ref{fig3}.}
\label{fig4}
\end{figure} 

The differences between the $z=1$ and $z=2$ results are largely
quantitative but once again show the influence of the anisotropy of
the qp scattering. The $B_{1g}$ spectra show the greater difference between
the two scaling regimes, as the overall spectra appears flatter for
$z=1$ than for $z=2$. This is due to the more pronounced hot spots in the 
$z=1$ regime. The $B_{1g}$ channel thus is more sensitive than the $B_{2g}$ 
channel to the growth of the correlation length as temperature is lowered.
Features of a crossover from $z=2$ to $z=1$ scaling as the temperature
is lowered through $T_{cr}$ would be most evident in the $B_{1g}$ 
channel\cite{la214}.

The role of the effective coupling constant is to determine the overall
shape of the continuum due to the increase in the qp scattering rate. An 
increase of the coupling constant leads to (1) an increase in the overall
magnitude of the cross section, (2) flatter spectra, and (3) the
movement of the weak peak of the spectra out to larger energy scales. 
This is shown in Fig. \ref{fig4} which plots the 
$B_{1g}$ and $B_{2g}$ response
as a function of frequency shift at low temperatures
$T/2t=0.03$ for different values of $g_{eff}=\alpha g^{2}/2t$.

As a consequence, the sensitivity of the Raman spectra to the details of the 
parameters can be a very useful tool to probe the anisotropic qp scattering
rates in much the same way as it has been used to probe the anisotropy
of the energy gap $\Delta({\bf k})$ in the superconducting state\cite{ijmpb}.

\section{Fit to overdoped Bi-2212.}

As we remarked, the approximations used ($i-iv$ listed in Section III B) 
are most appropriate for 
systems with weak spin fluctuation scattering and therefore we expect that 
our results would best represent the data from appreciably over-doped cuprate 
superconductors. Before we consider the effects of a more consistent
treatment of the interactions in NAFL beyond perturbation theory, it is
useful to explore how the perturbative theory compares to the data on
overdoped cuprates in their normal state.

For a comparison to these data we thus consider the results 
from the $z=2$ scaling regime in closer detail.
Over-doped samples typically have higher residual resistivities than
optimally doped samples. While this may be due in part to sample
preparation, it is believed that 
the over-doped cuprates increased effective dimensionality 
allows the qps to interact more strongly with defects residing
out of the CuO$_{2}$ planes\cite{ijmpb}. 
Therefore in addition we will consider
an isotropic impurity interaction 
$H_{imp}=\sum_{\bf k,k^{\prime}}\sum_{i,\sigma}
U e^{i({\bf k-k^{\prime}})\cdot {\bf R}_{i}}
c^{\dagger}_{{\bf k},\sigma}c_{{\bf k^{\prime}},\sigma},$
where ${\bf R}_{i}$ denotes the position of the impurity labeled by
$i$ and $U$ is the impurity potential. After averaging over the position of the
impurities, this adds 
a momentum independent term to the imaginary part of the self energy 
$\Gamma_{imp}=\pi n_{i}N_{F}\mid U\mid^{2}$, where $n_{i}$ is
the impurity concentration and $N_{F}$ is the density of states per spin
at the Fermi level.

\begin{figure}
\psfig{file=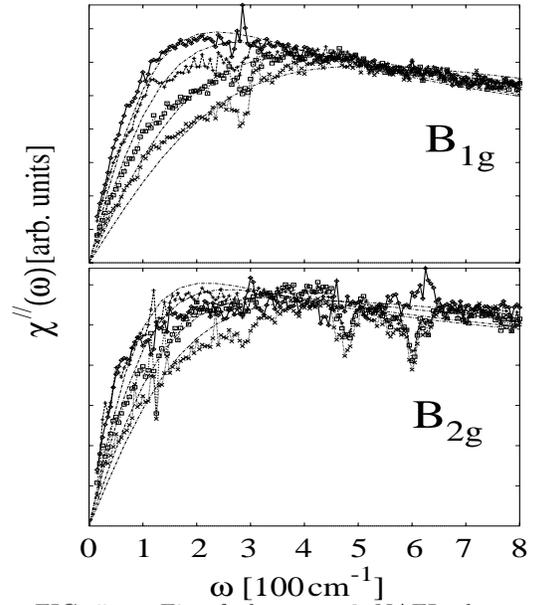,height=8.cm,width=7.cm,angle=0}
\caption[]{
Fit of the $z=2$ NAFL theory to the $B_{1g}$ and $B_{2g}$ spectra taken in 
\cite{hackl} on appreciably 
over-doped Bi$_{2}$Sr$_{2}$CaCu$_{2}$O$_{8+\delta}$ 
(T$_{c}=55K$) at $60K, 90K, 150K,$ and $200K$ from top to bottom. Parameters 
used are given in the text.}
\label{fig5}
%\vskip -0.5cm
\end{figure} 

The fit of the theory to the temperature dependent spectra for each channel 
obtained in appreciably
over-doped Bi 2212 (T$_{c}=55 K$) by Hackl et al. in Ref. 
\cite{hackl} is shown in Fig. \ref{fig5} 
for the $B_{1g}$ and $B_{2g}$ channels.
The impurity scattering rate for this material can be estimated from the
extrapolated $T=0$ dc resistivity of approximately
$30 \mu \Omega -$cm\cite{Kend93}. 
Assuming a Drude resistivity and using a plasma frequency of $1.6$ eV given
in \cite{Kend93} yields $\Gamma_{imp}=40$ cm$^{-1}$. 
We note that 
$\omega_{sf}$ and $\xi$ have so far not been determined via fits to magnetic
response data. The fits are obtained by choosing: $\omega_{sf}$ and 
$\xi$ at a fixed temperature, 
a magnitude of the coupling constant $\alpha g^{2}$, and lastly $b_{1}$
and $b_{2}$. Then, only the temperature dependent part of
$\omega_{sf}$ was modified to fit the data at other temperatures. 
We have used $b_{2}/b_{1}=0.417, 
\omega_{sf}\xi^{2}=13$meV, and $\omega_{sf}=160K+0.06T[K]$. This parameter
choice leaks to remarkably good agreement with the data. The flatness of the
continuum for both channels emerges from the theory and the relative peak
positions come out correctly.  More importantly, for the first
time the temperature dependence of the spectra in both 
channels can be explained. NAFL provides an excellent
description of the qp dynamics in the over-doped region.

Further information on qp dynamics
can be obtained from the slope of the spectra
at vanishing frequencies. We denote this slope as
$\Gamma_{\lambda}=\lim_{\Omega \rightarrow 0} \Omega/
\chi^{\prime\prime}_{\lambda}(\Omega)$, for $\lambda=B_{1g}$ or
$B_{2g}$. In Ref. \cite{hackl} it was shown that the
inverse of this slope has qualitatively different behaviors for different
doping regimes of various cuprate materials. While the inverse slope
observed in the $B_{2g}$ channel always was found to track the temperature
dependence of the DC resistivity for all regions of doping, the $B_{1g}$
inverse slope showed a remarkable sensitivity to doping. For overdoped
systems, the $B_{1g}$ slope was similar to the $B_{2g}$ rate, and
effectively followed the behavior $\Gamma_{B_{1g}}
\sim \Gamma_{B_{2g}} \sim \rm{constant} +T^{2}$. For
optimally doped cuprates, $\Gamma_{B_{2g}} \sim T$, while
$\Gamma_{B_{1g}} \sim$ constant. Moreover, for underdoped materials
the $B_{1g}$ inverse slope was found to increase with decreasing
temperature, indicative of insulating behavior.

Within the level of our
approximations, the low frequency Raman response is given by
\begin{eqnarray}
&&1/\Gamma_{\gamma}(T)=\lim_{\Omega \rightarrow 0}
\chi^{\prime\prime}_{\gamma\gamma}(\Omega)/\Omega= 
\label{fifteen}
\\
&&\int
{{\rm d}\Omega_{\bf k}\over{\mid {\bf v_{k}}\mid}}\gamma^{2}({\bf k})\int
{{\rm d}x\over{[2\cosh(x/2)]^{2}}}{1\over{\Sigma^{\prime\prime}({\bf k},xT)}}.
\nonumber
\end{eqnarray}
The inverse slope samples
the qp scattering lifetime at regions of the Fermi surface selected by
the light polarization orientations. It is important to note that
in this level of approximation the momentum dependence of the Raman
vertices competes with the momentum dependent scattering rate on the
Fermi surface since the largest contribution to the integral
in Eq. (\ref{fifteen}) comes from where the Raman vertex is the largest and
where the qp scattering rate is the smallest. Therefore, the Raman
inverse slope washes out direct information about the $k-$
dependence of the scattering rate and leads to a more similar
behavior of the $B_{1g}$ and $B_{2g}$ inverse slope, which hides
the underlying anisotropy of the qp scattering rate. This is due to 
the approximations of restricting consideration to the Fermi surface
only and to the neglect of vertex corrections. In the following section 
this is shown to be crucial if the spin fluctuations are strong.

\begin{figure}
\psfig{file=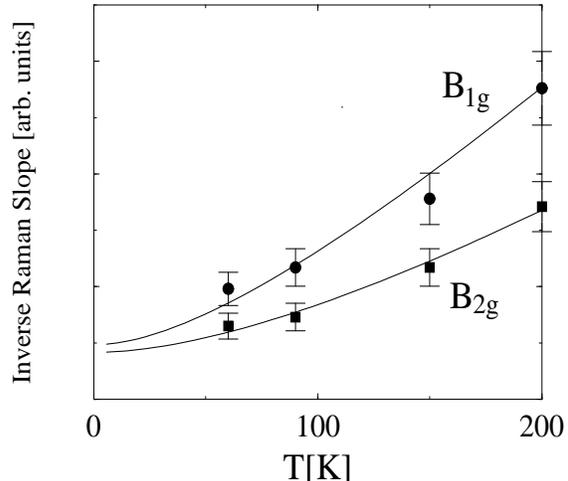,height=7.cm,width=8.cm,angle=0}
\caption[]{
Fit of the inverse slope of the Raman response for the two symmetry
channels. The error bars are from Ref. \cite{opel}.}
\label{fig6}
\end{figure} 

However, for weak spin fluctuations, we believe that the data on overdoped
materials can be best fit with this perturbative approach.
In Fig. \ref{fig6} we plot the inverse Raman slope obtained 
from the parameters used in Fig. \ref{fig5} and compare the results to the 
data taken in \cite{hackl}. The agreement confirms the fits in Fig. 
(\ref{fig5}). The data are consistent with
a scattering rate which is not too anisotropic, and is governed by
elastic scattering from impurities and Fermi liquid inelastic 
scattering from spin fluctuations. We remark that the fits
are robust to small NAFL parameter changes.

\section{Self consistency and vertex corrections.}

While good agreement between the theory and the data has been shown for 
over-doped Bi 2212, several features remain
to be explained in optimally and under-doped systems. As mentioned in
the last section,
the difference of the inverse slope of the 
Raman response between $B_{1g}$ and $B_{2g}$ channels grows remarkably
upon lesser dopings\cite{hackl}. Moreover, features which have been
associated with a pseudo-gap have been observed in the under-doped 
systems\cite{psuedo,psuedo2}.  Since our fits to the data were restricted by
our approximations to weak spin fluctuations, we need an improved
formalism for an insight into the effects of strong spin fluctuations.

Therefore for the remainder of the letter we solve the model lifting the 
restrictions $(i-iv)$ listed before.
The calculations were performed by first solving for the Green's function
Eq. (\ref{fourteen}) self consistently while maintaining a filling 
$\langle n \rangle=0.8$. We used a hard frequency cutoff $\omega^{*}=0.4$eV 
for the interaction Eq. (1) to improve convergence, but note that our results 
hardly were affected by this cutoff. Then the Bethe-Salpeter equation 
(\ref{ten}) was evaluated for the renormalized vertex. 
The calculations 
were carried out on the imaginary frequency axis using 256 Matsubara
frequencies and $64\times 64$ to $128\times 128$ ${\bf k}$-points. 
These values are similar to that used for resistivity studies\cite{MP}.
The integral
equation for the vertex $\tilde\gamma({\bf k},i\omega,i\omega+i\Omega)$ was 
solved for each external (internal) bosonic (fermionic)
Matsubara frequency $i\Omega$ ($i\omega$). Once completed,
the Raman response was evaluated using Eq. \ref{nine}
and the results were analytically continued
to the real frequency axis using Pad\'e approximants\cite{pade}. Since the 
Raman response is in general quite featureless the Pad\'e procedure was highly
accurate (using 20-60 points changes the results by less than 5 \%).

\begin{figure}
\psfig{file=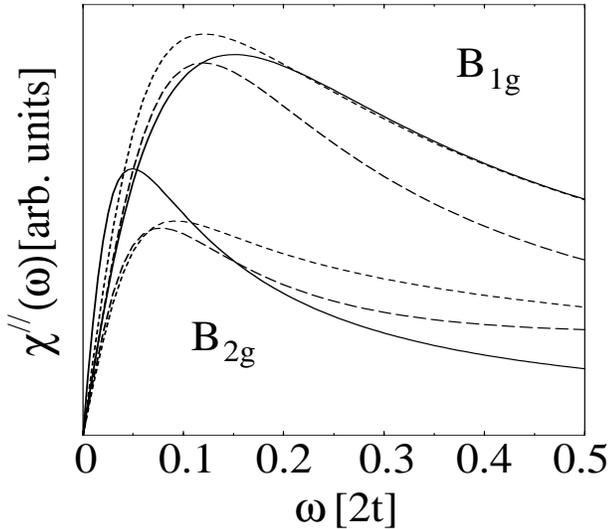,height=7.cm,width=8.cm,angle=0}
\caption[]{Comparison of the $z=2$ (over-doped)
results obtained from perturbation theory (solid line),
self consistent treatment (short dashed line), and the additional 
inclusion of vertex corrections (long dashed lines).}
\label{fig7}
%\vskip -0.5cm
\end{figure} 

We first consider how the perturbation results for the $z=2$ regime
are affected.
The results are summarized in Fig. \ref{fig7} which compares the effects of
self consistency and vertex renormalizations to the results from 
perturbation theory for $T=0.08t$. $\omega_{sf},\xi,$ and $g_{eff}$
were chosen as in Fig. \ref{fig3}.
By noting that the position of the peak in the $B_{1g}$ channel decreases
in frequency while the $B_{2g}$ peak position increases under self
consistency, it can be seen
that the effect of a self consistent treatment is to decrease the relative
anisotropy in the scattering rate around the Fermi surface leading to
a more effectively ``isotropic'' response which places the peak of
the spectra at nearly the same position $(\sim 0.2t)$ for both channels.
The $B_{1g}$ spectrum
is slightly narrower in profile than the perturbation results, while
the $B_{2g}$ spectrum is wider. Although we
have not tried to fit the self consistent theory to the data on
over doped Bi 2212 in Fig. \ref{fig5}, it appears that if disorder is
once again added, an equally good fit would be obtained using a slightly
larger value of the coupling constant $g_{eff}$ to offset the reduction
due to self consistency. We note however that the $z=2$ self consistent
treatment would not yield a particularly good fit to the data on
optimally and underdoped systems, which show much more anisotropy than
that indicated in Fig. \ref{fig7}.
 
\begin{figure}
\psfig{file=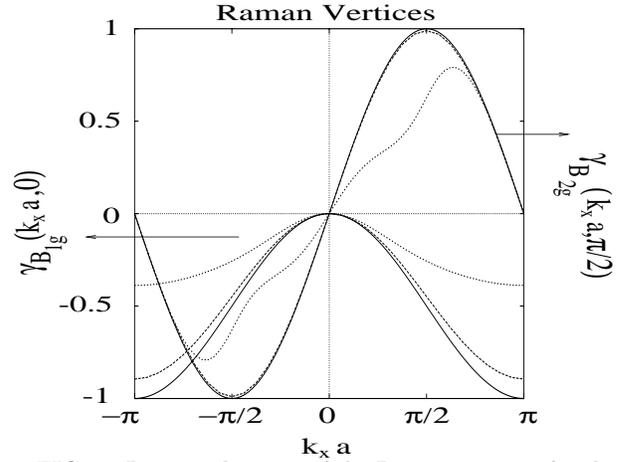,height=6.cm,width=8.cm,angle=0}
\caption[]{Renormalization of the Raman vertices
for the two symmetry channels evaluated for the lowest bosonic
Matsubara frequency. The solid lines are the bare vertices, the dashed
are 
the renormalized vertices for the $z=2$ (over-doped) parameters as in
Fig. \ref{fig3}, and the dotted lines are the  
renormalized vertices corresponding to $z=1$ (under-doped) parameters,
using $\alpha=16.7$ states/eV, corresponding to YBa$_{2}$Cu$_{3}$O$_{6.5}$
\cite{rmp}.}
\label{fig8}
\end{figure} 

This can be partially offset by interaction renormalization of the
Raman vertex. When the vertex corrections are included, the $B_{1g}$
spectrum is most affected, with an overall reduction intensity occurring
over a wide region of frequencies, which we interpret as effects
related to precursor SDW formation\cite{rmp}. 
This can be seen more clearly by examining the Raman vertices directly.
In Fig. \ref{fig8} we plot the $B_{1g}$ and $B_{2g}$ vertices for selected 
directions in the Brillouin Zone for different values of the coupling 
$g_{eff}$ and different scaling regimes. The $z=2$ curves correspond to the 
parameters used in Figs. \ref{fig3} and \ref{fig7}, 
while the $z=1$ curves correspond to 
the values used for YBa$_{2}$Cu$_{3}$O$_{6.5}$ in \cite{rmp}, with 
$\alpha=16.7$ states/eV. Since $g=1$ eV and $t=250$ meV 
has been used for all curves, $g_{eff}=\alpha g^{2}/2t=6.2 (33.4)$ for the
$z=2 (z=1)$ curves in Fig. \ref{fig8}.
We found no cases of symmetry mixing and the
renormalized vertices retain their bare group transformation properties.
Compared to the bare vertices, the $B_{1g}$ vertex shows much more of
reduction than the $B_{2g}$ vertex. The renormalized vertex 
$\tilde\gamma$ for the $B_{1g}$ channel
is reduced in particular near the zone axes or ``hot spots''.
However, the $B_{2g}$ vertex is hardly affected by the vertex 
renormalizations. This is consistent with the effects of a ``cold''
interaction probed along the zone diagonals. For $z=2$ the effects of
the vertex corrections are small but increase dramatically for $z=1$.
As the interaction is
increased, the $B_{1g}$ vertex is reduced dramatically near the ``hot
spots''. For $z=1$ the 
$B_{1g}$ vertex is reduced by a factor 2 near the ``hot spots'' 
compared to its bare value, while the $B_{2g}$ vertex is reduced by a 
smaller value $ 25 \% $, and has a peak which is shifted out towards 
the ``hot spots''. Since the Raman vertex renormalization
enters with the opposite sign as the self energy vertex correction, these 
results complement the results of Ref. \cite{chubukov} which showed that the 
effect of vertex corrections {\it on the interaction} yields an effectively 
increased coupling constant. 

The vertex renormalization suppress the $B_{1g}$ cross
section compared to the $B_{2g}$ cross section for larger interaction
strengths. Since the $B_{2g}$ vertex is affected less dramatically, the
net effect would appear that the $B_{2g}$ cross section remains
relatively unchanged with increasing interaction, while the $B_{1g}$
overall cross section comes down in magnitude to be closer to the
$B_{2g}$ value. We remark that this trend is consistent with the 
data from the cuprates which show a large decrease of the $B_{1g}$ to $B_{2g}$ 
ratio of the Raman spectra upon under-doping\cite{psuedo} which we relate to 
the growth of ``hot spots'' intensity and subsequently large vertex
renormalizations.

\begin{figure}
\psfig{file=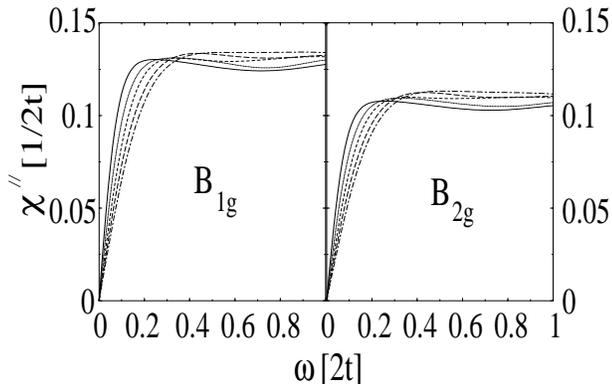,height=5.cm,width=8.cm,angle=0}
\caption[]{The full Raman response for $z=1$ (under-doped)
(evaluated with self consistency
and vertex renormalization) at temperatures $T/2t=0.03$ (solid),
0.04 (dotted), 0.5 (dashed), 0.6 (long-dashed), 0.7 (short-dashed),
for the $B_{1g}$ and $B_{2g}$ channels. The parameters used
corresponding to values for YBa$_{2}$Cu$_{3}$O$_{6.5}$\cite{rmp}.}
\label{fig9}
\end{figure} 

This can be seen more clearly in Fig. \ref{fig9}, which shows the full
Raman response for
larger values of the coupling $g_{eff}$ corresponding to 
YBa$_{2}$Cu$_{3}$O$_{6.5}$\cite{rmp}. The spectra are much flatter
than those calculated for smaller coupling constants, and appear to
differ only quantitatively. The effect of the anisotropy of the scattering 
rate appears to have been washed out due to self consistency, as the
spectra have a ``shoulder'' at the same frequency and show roughly the
same temperature dependence. However, the overall
magnitude of the spectra at larger frequency shifts is reduced for the
$B_{1g}$ spectra relative to the $B_{2g}$ spectra, leading to a ratio
of the continuum nearly equal to 1. As indicated in Fig. \ref{fig8}, this
is entirely due to the vertex renormalization. While the relative
magnitude of the continua and overall
lineshape is similar to that seen in experiments at a single temperature
\cite{psuedo}, the temperature dependence seen in experiments is
qualitatively different.

\begin{figure}
\psfig{file=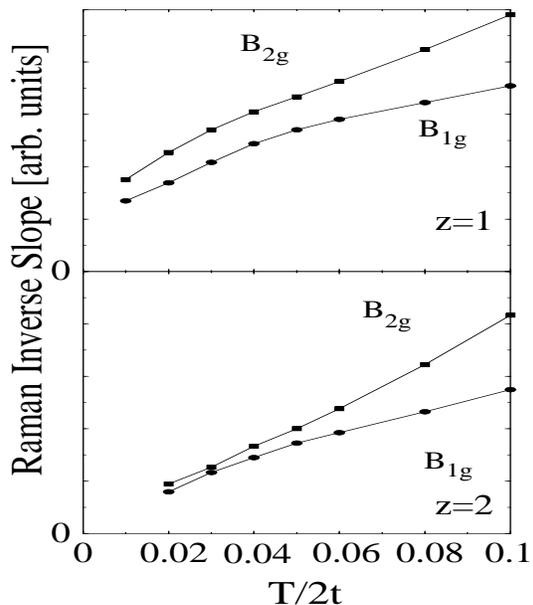,height=8.cm,width=7.cm,angle=0}
\caption[]{The temperature dependence of the Raman inverse slope 
calculated self consistently with vertex corrections
evaluated for both symmetries and scaling regimes.} 
\label{fig10}
%\vskip -0.5cm
\end{figure} 

This can be explored in more detail by plotting in 
Fig. \ref{fig10} the inverse slope 
of the Raman spectra for the self consistent and vertex corrected calculation.
These results are qualitatively different from the perturbative
treatment as discussed earlier. From Eq. (\ref{fifteen})
the inverse Raman slope has the same behavior 
as the qp scattering rate at zero frequency. If no disorder is included then
the inverse Raman slope varies as $T^{2}$ below a characteristic $T^{*}$ and
crosses over to $T$ at larger temperatures. $T^{*}$ is connected with
$\omega_{sf}$, and is generally dependent on parameter choices. However
Fig. \ref{fig10} shows that the vertex corrected, self consistent treatment 
does
not show similar behavior. For the z=2 scaling regime, the inverse slope
is proportional to $T$ down to the lowest temperatures reachable for our
level of numerical accuracy, while for the $z=1$ regime, the inverse slope 
appears to extrapolate to a non-zero value at $T=0$. This is also seen
in FLEX calculations\cite{altmann}.
A simple connection between
these and the perturbative results is that the self consistency leads
to a strong downward renormalization of $\omega_{sf}$ and a much lower
crossover $T^{*}$ to Fermi liquid-like behavior. The renormalization is 
stronger
for the $z=1$ case than $z=2$, in agreement with our earlier discussion.
Our results thus indicate that
the inverse slope should fall as $T$ is decreased until $\sim T_{cr}$
which separates the $z=1$ and $z=2$ scaling behavior, at which point
the slope should become less $T-$dependent. Such
behavior has been reported by Naeini {\it et al.} in 
slightly overdoped La$_{1-x}$Sr$_{x}$CuO$_{4}$\cite{la214} and agrees
well with the NAFL results.

While this behavior is similar to what has been observed in the
$B_{2g}$ channel, the theory does not predict a considerable difference
for the $B_{1g}$ channel. 
From our calculations, it appears that the effect of
self consistency obscures the anisotropy inherent in the model, leading
to a temperature dependent behavior which is similar for both channels.
While the effect of vertex corrections appears to heighten the anisotropy,
as shown in Fig. \ref{fig8}, the self consistency appears to dominate the
response, as seen in Fig. \ref{fig10}. It would appear then that a
treatment which neglected self consistency and focused only on vertex
renormalizations would lead to a more apt description of the experiments.
This remains to be explored.
 
\section{summary and conclusions}

In summary we have seen how anisotropic qp scattering leads to manifestly
channel dependent Raman cross sections and how through a careful examination
of the temperature dependence of the spectra, important information can be
extracted concerning anisotropic qp dynamics. The coherence or incoherence
of the qp's spectral function can be probed directly via Raman scattering,
thereby offering an additional avenue besides ARPES
to explore momentum resolved qp information. Signatures of a strong
interaction are: (1) the position of a weak 
peak in the spectrum occurring at large frequency shifts, (2)
the increased overall magnitude of the spectra, and (3) anomalous
temperature dependence of the spectra. If the interaction
is strongly anisotropic, these features are
different for different polarization channels. 
The qp dynamics can be probed by examining the temperature
dependence of both the low and high
frequency portions of the channel dependent Raman response. 
For the cuprates, the consequence of an interaction which
scatters qps with relative momentum of the antiferromagnetic lattice
vector ${\bf Q}$ are (1) a stronger overall signal, (2) a peak at higher 
frequencies, and (3) a weaker temperature dependence of the slope of
the response in the $B_{1g}$ channel compared to the $B_{2g}$ channel.

Excellent agreement of the NAFL model was found with 
the Raman results on overdoped Bi 2212 (and La 214 \cite{la214}).
With the inclusion of impurity scattering, the Raman response 
calculated in a perturbative way in NAFL yields a correct
description of the spectra at a variety of temperatures.
Improvements to the theory
can be made with more accurate determination of the parameters of the
theory via other experimental probes.

The effect of a more self consistent treatment leads to an overall
reduction of anisotropy effects and a more isotropic Raman response,
in contradiction to the experiments on optimally- and under-doped
cuprates, at least for the $B_{1g}$ channel. The
self consistency yields a much weaker Fermi liquid-like behavior than
that obtained via a perturbative approach. However,
the theory does not
adequately describe the larger relatively incoherent scattering probed
by the $B_{1g}$ light orientations. Improvements can be made by once
again exploring different parameter choices, but the nearly temperature
independence of the $B_{1g}$ response seen in the cuprates must be obtained
by tuning both $\omega_{sf}$ and $\xi$ to have a strongly increasing
interaction peaked near $(\pi,\pi)$ transfers to offset the loss of
phase space for qp scattering as the temperature is lowered. This remains
to be explored.

Interestingly, the role of vertex corrections may help to obtain such a
behavior. For larger couplings, the 
vertex corrections were found to be particularly large
near the ``hot spots'' and led to a subsequently reduced vertex
and overall signal for $B_{1g}$ polarizations. Only a small
change was seen for the $B_{2g}$ channel. This general behavior
has been observed in Y 123 and La 214, and has been attributed to the
opening of a pseudo-gap. More than likely it would be important to 
also employ vertex corrections to the self energy and to go beyond the
simple calculations as performed here. 
Diagrammatically we have evaluated the
lowest order diagrams for the self energy, and we include the interactions
in the Raman response through uncrossed vertex corrections (ladder
approximation). We do not expect that these calculations can describe very
strong spin fluctuations, but our approach in principle should give
us an insight as to what is the most important physics we might expect
as the strength of the spin fluctuations is increased by removing
oxygen. This remains an open challenge.

\acknowledgements
The authors would like to thank M. Opel, R. Nemetschek, and
R. Hackl for making the data
presented in Fig. \ref{fig5} available to us prior to publication.
T.P.D. would like to acknowledge helpful conversations with D. Pines, J.
Schmalian, and
B. Stojkovi\'c. Acknowledgment (T.P.D.) is made to the Donors of the Petroleum
Research Fund, administered by the American Chemical Society, for
partial support of this research.


\vskip -0.5cm
\begin{references} 
\bibitem{marshall} D.S. Marshall, D. S. Dessau, A. G. Loeser, C.-H. Park,
A. Y. Matsuura, J. N. Eckstein, I. Bozovic, P. Fournier, A. Kapitulnik,
W. E. Spicer, and Z.-X. Shen, Phys. Rev. Lett. {\bf 76}, 4841 (1996).

\bibitem{pgap}
A. G. Loeser, Z.-X. Shen, D. S. Dessau, D. S. Marshall, C. H. Park, P.
Fournier, and A. Kapitulnik, Science {\bf 273}, 325 (1996);
H. Ding, M. R. Norman, T. Yokoya, T. Takeuchi, M. Randeria, J. C.
Campuzano, T. Takahashi, T. Mochiku, and K. Kadowaki, Phys. Rev.
Lett. {\bf 78}, 2628 (1997).

\bibitem{ijmpb} 
T. P. Devereaux, D. Einzel, B. Stadlober, R. Hackl,
D. H. Leach, and J. J. Neumeier, Phys. Rev. Lett. {\bf 72}, 396 (1994).
For a recent review, see
T. P. Devereaux and A. P. Kampf, Int. J. Mod. Phys. B {\bf 11}, 2093 (1997).

\bibitem{ubcdata} 
G. Blumberg, P. Abbamonte, M. V. Klein, W. C. Lee, D. M. 
Ginsberg, L. L. Miller, and A. Zibold, Phys. Rev. B {\bf 53}, R11930
(1996); D. Einzel and R. Hackl, J. Raman Spectroscopy {\bf 27}, 307 (1996);
D. Reznik, S. L. Cooper, M. V. Klein, W. C. Lee, D. M. 
Ginsberg, A. A. Maksimov, A. V. Puchkov, I. I. Tartakovskii, and S.-W.
Cheong, Phys. Rev. B {\bf 48}, 7624 (1993).

\bibitem{rhoraman} 
R. Hackl, M. Opel, P. F. M\"uller, G. Krug, B. Stadlober, R. 
Nemetschek, H. Berger, and L. Forr\'o, J. Low Temp. Phys. {\bf 105},
733 (1996).

\bibitem{hackl}
R. Hackl, G. Krug, R. Nemetschek, M. Opel, and B. Stadlober, in {\it
Spectroscopic Studies of Superconductors}, I. Bozovic and D. van der Marel,
Eds., Proc. SPIE 2696, 194 (1996).

\bibitem{psuedo}
X. K. Chen, J. G. Naeini, K. C. Hewitt, J. C. Irwin, R. Liang, and
W. N. Hardy, Phys. Rev. B {\bf 56}, R513 (1997); 
R. Nemetschek, M. Opel, C. Hoffmann, P. F. M\"uller, R. Hackl, H.
Berger, L. Forr\'o, A. Erb, and E. Walker, 
Phys. Rev. Lett. {\bf 78}, 4837 (1997). 

\bibitem{psuedo2}
A small temperature dependent peak at 600 cm$^{-1}$ in the $B_{1g}$ channel
has recently been observed [G. Blumberg, M. Kang, M. V. 
Klein, K. Kadowaki, and C. Kendziora, Science, {\bf 278}, 1427 (1997)]. 

\bibitem{im}
L. B. Ioffe and A. J. Millis, Phys. Rev. B {\bf 58},
11631 (1998).

\bibitem{zxrs}
Z.-X. Shen and J. R. Schrieffer, Phys. Rev. Lett. {\bf 78}, 1771 (1997).

\bibitem{altmann} 
J. Altmann, W. Brenig, and A. P. Kampf, preprint cond-mat 9707267.

\bibitem{MMP}
A. Millis, H. Monien, and D. Pines, Phys. Rev. B {\bf 42},167 (1990).

\bibitem{rmp}
D. Pines, Z. Phys. B {\bf 103}, 129 (1997);
A. V. Chubukov, D. Pines, and B. Stojkovi\'c, J. Phys. Cond. Matt.
{\bf 8}, 10017 (1996); J. Schmalian, D. Pines, and B. Stojkovi\'c,
cond-mat/9804129 and references therein.

\bibitem{ps}
D. Pines and B. Stojkovi\'c, Phys. Rev. B {\bf 55}, 8576 (1997);
{\bf 56}, 11931 (1997).

\bibitem{kandd}
M. V. Klein and S. B. Dierker, Phys. Rev. B {\bf 29}, 4976 (1984);
H. Monien and A. Zawadowski, Phys. Rev. B {\bf 41}, 8798 (1990).

\bibitem{kz}
J. Kosztin and A. Zawadowski, Solid State Commun. {\bf 78}, 1029 (1991).

\bibitem{ag}
A. A. Abrikosov and V. M. Genkin, Zh. Eksp. Teor. Fiz. {\bf 40}, 842 (1973) 
[Sov. Phys. JETP {\bf 38}, 417 (1974)].

\bibitem{allen}
P. Allen, Phys. Rev. B {\bf 13}, 1416 (1976).

\bibitem{zawa}
A. Zawadowski and A. Cardona, Phys. Rev. B {\bf 42}, 10732 (1990).

\bibitem{nfl}
A. Virosztek and J. Ruvalds, Phys. Rev. B {\bf 45}, 347 (1992).

\bibitem{chub}
A. V. Chubukov, D. Pines, and B. P. Stojkovi\'c, J. Phys. Condens.
Matter {\bf 8}, 10017 (1996).

\bibitem{bulut}
N. Bulut, D. Scalapino, and S. R. White, J. Chem. Phys. Solids
{\bf 54}, 1109 (1993).

\bibitem{la214}
J. G. Naeini, X. K. Chen, K. C. Hewitt, J. C. Irwin, T. P. Devereaux,
M. Okuya, T. Kimura, and K. Kishio, Phys. Rev. B {\bf 57},
R11077 (1998).

\bibitem{Kend93}
C. Kendziora, M. C. Martin, J. Hartge, L. Mihaly, and L. Forr\'o,
Phys. Rev. B {\bf 48}, 3531 (1993).

\bibitem{opel}
M. Opel, private communication.

\bibitem{MP}
P. Monthoux and D. Pines, Phys. Rev. B {\bf 49}, 4261 (1994).

\bibitem{pade}
J. Vidberg and J. Serene, J. of Low Temp. Phys. {\bf 29}, 179 (1977).

\bibitem{chubukov}
A. V. Chubukov. P. Monthoux, and D. Moor, 
Phys. Rev. B {\bf 56}, 7789 (1997); J. Schmalian, private communication.

\end{references}
\end{document}